\documentclass[twocolumn,pra,superscriptaddress]{revtex4-2}
\usepackage{amsmath}
\usepackage{amssymb}
\usepackage{graphicx}
\usepackage{bm}
\usepackage{booktabs}
\usepackage{multirow}
\usepackage{color}

\makeatletter

\begin{document}

\title{Two-parameter estimation with three-mode NOON state in a symmetric triple-well system}

\author{Fei Yao}
\affiliation{Graduate School of China Academy of Engineering Physics, Beijing 100193, China}

\author{Yi-Mu Du}
\affiliation{Graduate School of China Academy of Engineering Physics, Beijing 100193, China}

\author{Haijun Xing}
\email{hjxing3@icloud.com}
\affiliation{Graduate School of China Academy of Engineering Physics, Beijing 100193, China}

\author{Libin Fu}
\email{lbfu@gscaep.ac.cn}
\affiliation{Graduate School of China Academy of Engineering Physics, Beijing 100193, China}

\begin{abstract}
We propose a scheme to realize two-parameter estimation via a Bose-Einstein condensates confined in a symmetric triple-well potential. The three-mode NOON state is prepared adiabatically as the initial state. The two parameters to be estimated are the phase differences between the wells.
The sensitivity of this estimation scheme is studied by comparing quantum and classical Fisher information matrices.  As a result, we find an optimal particle number measurement method. Moreover, the precision of this estimation scheme behaves the Heisenberg scaling under the optimal measurement.
\end{abstract}

\maketitle

\section{Introduction}

Quantum metrology \cite{GLM2004,GLM2006,GLM2011} has attracted considerable interest in recent years due to its wide applications in both fundamental sciences and applied technologies. As crucial tools in quantum metrology, the quantum parameter estimation theory~\cite{Helstrom1976,Holevo1982} and Fisher information provide the theoretical bases for enhancing the precision of parameter estimation with quantum resources.
In the previous researches, 
the single-parameter estimation has been well studied and a series of achievements have been made~\cite{Pezze2018,Degen2017,Braun2018},  such as gravitational wave detection~\cite{LIGO2011}, magnetometry~\cite{Sewell2012,Ockeloen2013,Muesselll2014}, atomic clocks~\cite{Kruse2016,Pedrozo2020}, and quantum gyroscope~\cite{Stevenson2015,Che2018,Yao2018}.

Although the single parameter estimation plays a significant role in many applications, it is often necessary to estimate multiple parameters simultaneously in practical problems, e.g., quantum imaging \cite{Genovese2016,Rehacek2017,Ang2017}, waveform estimation \cite{Tsang2011},
 measurements of multidimensional fields~\cite{Baumgratz2016}, joint estimation of phase and phase diffusion~\cite{Vidrighin2014,Altorio2015}.
Studying the multi-parameter estimation is thus an urgent need for effectively solving the practical  parameter estimation problems. It have attracted lots of attentions~\cite{Vidrighin2014,Altorio2015,Crowley2014,Kok2017,Humphreys2013,Szczykulska2016,Ragy2016,Baumgratz2016,Pezz2017,Zhuang2018,Gessner2018,Gessner2020,Goldberg2020,Goldberg2021,Xing2020,Liu2020,Lu2021} in recent years. Most of these works aim to propose a general theory and framework for multi-parameter estimation. Few concrete schemes are proposed for realizing  practical high-precision multi-parameter estimations. In this article, we will propose a scheme to estimate multiple parameters simultaneously with Heisenberg scaling sensitivity.

Bose-Josephson junction, formed by confining Bose-Einstein condensate in the double-well potential (in spatial freedom~\cite{Smerzi1997} or internal freedom~\cite{Cirac1998,Yi2006} equivalently) is a well-established model \cite{Leggett2001,Cronin2009,Pezze2018,Lee2012}.
It is widely used in quantum parameter estimation as interferometries for its high controllability~\cite{Leggett2001,Cronin2009,Pezze2018,Lee2012}. Especially in some of the schemes \cite{Yi2006,Bychek2018,Pezz2019,Lee2006,Xing2016}, one can prepare condensate into the two-mode NOON state \cite{Lee2002} (also known as GHZ state \cite{GHZ1990} and Schr\"{o}dinger cat state \cite{Bollinger1996}), which can preform single parameter estimation in Heisenberg limit precision.  As an extension of the double-well interferometry, we will confine Bose condensate in the symmetrical triple-well potential~\cite{Nemoto2000,Franzosi2001,Franzosi2003,Cao2014,Cao2015} to realize high precision two-parameter estimation.

Our measurement scheme consists of four stages: initialization, parameterization, rotation, and measurement.
We prepare the condensate into the three-mode NOON state adiabatically as the initial state.
The parameters to be estimated are two phase differences  between the wells  caused by the external field.
The parameterized state is read via the particle number measurement.
In order to study the precision of the measurement scheme,
we calculate the quantum Fisher information matrix (QFIM) and classical Fisher information matrix (CFIM) on the two parameters. By comparing the CFIM and QFIM, we find that the measurement rotation time significantly affects the measurement precision, and the optimal rotation time is given.
In addition, the result shows that the optimal measurement precision of our scheme can approach the Heisenberg scaling.

The paper is organized as follows. In Sec.~\uppercase\expandafter{\romannumeral2}, the model and basic  measurement theory are introduced.
In Sec.~\uppercase\expandafter{\romannumeral3}, we give the scheme of estimating two parameters with the triple-well system, including initial state preparation, parameterization, rotation, and measurement. The optimal precision and measurement conditions are given by analyzing the CFIM and QFIM.
At last, we summarize this article in Sec.~\uppercase\expandafter{\romannumeral4}.

\section{Model and basic theory}

\begin{figure}[tp]
\includegraphics[width=75mm]{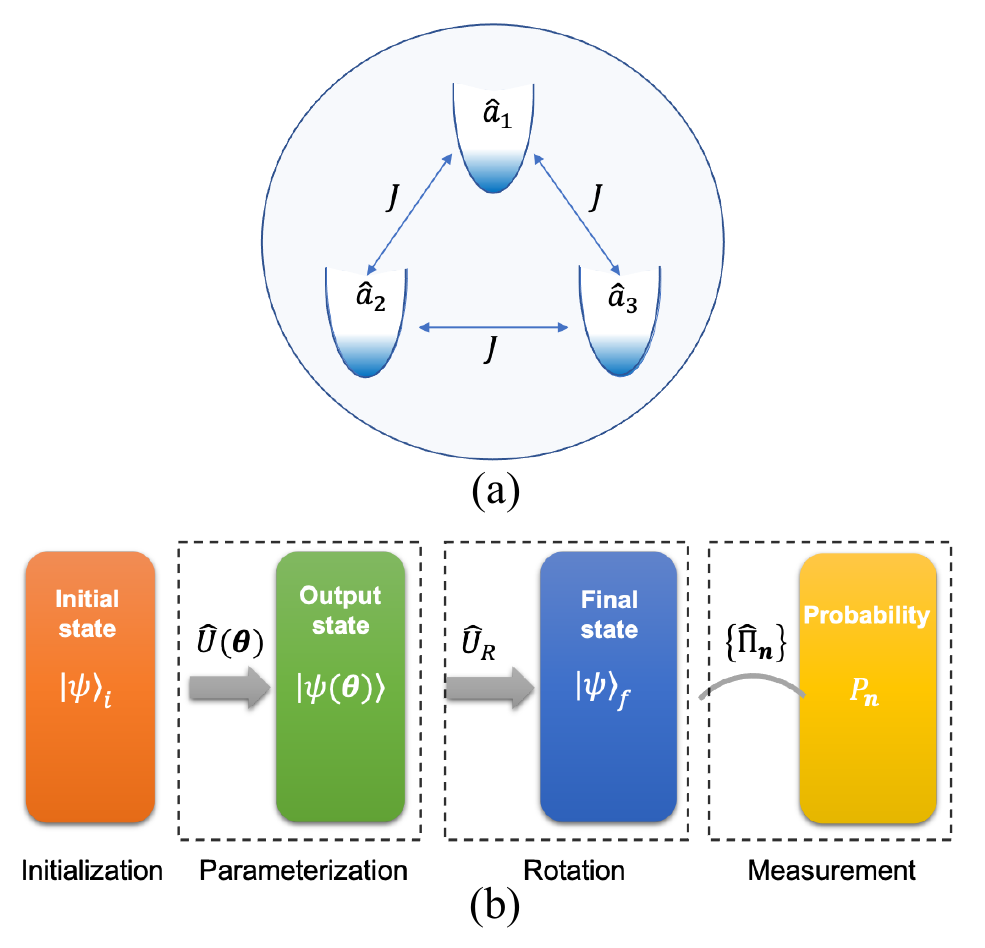}
\caption{(Color online) (a) The schematic diagram of a symmetric triple-well trapped Bose atoms.
(b) Framework of multi-parameter estimation.}
\label{Fig:model}
\end{figure}

In this section, we sketch our scheme. We confine $N$ Bose condensed atoms in a symmetric triple-well system(STWS)~\cite{Nemoto2000,Franzosi2001,Franzosi2003,Cao2014,Cao2015}, as seen in Fig.~\ref{Fig:model}(a).
Under the three-mode approximation \cite{Nemoto2000}, Hamiltonian of the condensates reads
\begin{equation}\label{eq:Hamiltonian}
\hat{H}=-J\sum_{i=1}^{3}(\hat{a}_{i}^{\dagger}\hat{a}_{j}+h.c.)+U\sum_{i=1}^{3}\hat{n}_{i}(\hat{n}_{i}-1),
\end{equation}
with $j=(i+1)$mod$3+1$.  The operator $\hat{a}_{i}^{\dagger}(\hat{a}_{i})$ is the bosonic creation (annihilation) operator for atoms in the ground state mode of $i$-th well, and $\hat{n}_i=\hat{a}_i^\dagger \hat{a}_i$ is the corresponding particle number operator. $J$ is the tunneling strength between the wells. It is controllable via adjusting the barriers between the wells.  $U$ is the atomic on-site interaction, and $U>0$~$(U<0)$ imply a repulsive~(attractive) interaction. It is tunable via the Feshbach resonances \cite{Chin2010}. Here, we only consider attractive interaction.
In this model, the total atom number $N=\sum_{i=1}^{3}n_{i}$ is conserved. The system state can be expanded on Fock state basis ${\{|n_{1},n_{2},n_{3}\rangle\}}$, with $n_i$ particle in the $i$-th well, $i=1,2,3$.

When this STWS is placed into an external field, the ground state energy of $i$-th well (mode $\hat{a}_i$)  experience an energy shift $\mathcal{E}_i$ denoted by the Hamiltonian
\begin{equation}
\hat{H}_p=\sum_i\mathcal{E}_i\hat{a}_i^\dagger\hat{a}_i,
\end{equation}
while the tunneling and interactions are both turned off ($J=U=0$).
Particle in mode $\hat{a}_i$ will obtain a phase shift $\phi_{i}=\mathcal{E}_i t$ after evolution generated by $\hat{H}_p$ with time $t$. Phase differences $\theta_{1}=\phi_{1}-\phi_{3}$ and $\theta_{2}=\phi_{2}-\phi_{3}$ are the parameters we aiming to estimate.

Before introducing details of our scheme, let us recall the framework of multi-parameter estimation as follows (see Fig.~\ref{Fig:model}(b)).
\begin{description}
\item [{i)}] Initialization: Prepare system to the initial state $|\psi\rangle_\mathfrak{i}$.
\item [{ii)}] Parameterization: The initial state $|\psi\rangle_\mathfrak{i}$  is parameterized to the output state  $|\psi(\bm\theta)\rangle=\hat{\mathcal{U}}(\bm\theta)|\psi\rangle_{\mathfrak{i}}$ via a unitary evolution $\hat{\mathcal{U}}({\bm{\theta}})$, where $\bm\theta=(\theta_1,\theta_2,\dots,\theta_d)$ is a vector parameter.
\item [{iii)}] Rotation: Rotate the output state $|\psi(\bm\theta)\rangle$ to the measurable final state
\begin{equation}
|\psi\rangle_\mathfrak{f}=\hat{\mathcal{U}}_{R}|\psi(\bm\theta)\rangle.
\label{eq:psif}
\end{equation} 
\item [{iv)}] Measurement: Perform a set of projective measurements $\{\hat{\Pi}_{\bm{n}}\}$ ($\bm{n}$ representing the possible result) on the final state $|\psi\rangle_\mathfrak{f}$. The probability of observing the result $\bm{n}$, which conditioned to the vector parameter $\bm{\theta}$, is
\begin{equation}
P_{\bm{n}}={}_{\mathfrak{f}}\langle\psi|\hat{\Pi}_{\bm{n}}|\psi\rangle_\mathfrak{f}.
\label{eq:p}
\end{equation}
\end{description}

Vector $\bm{\theta}$ is estimated based on $\{ P_{\boldsymbol{n}} \}$, statistics of the measurement results. In this article, we only discuss the unbiased estimation. Fisher information lies in  the heart of evaluating the precision of this estimation. 
For the probability shown in Eq.~\eqref{eq:p}, the matrix elements of CFIM $\bm{F}^{c}$ is defined as
\begin{equation}
\bm{F}_{\mu,\nu}^{c}=\sum_{\bm{n}}\frac{\partial_{\mu}P_{\bm{n}}\partial_{\nu}P_{\bm{n}}}{P_{\bm{n}}},
\label{eq:Fc}
\end{equation}
with $\partial_{\mu}:=\partial/\partial\theta_{\mu}$ and $\mu,\nu=1,2$. According to the quantum parameter estimation theory~\cite{Helstrom1976,Holevo1982}, $\boldsymbol{F}^c$ determines the best  precision of the unbiased estimators of $\boldsymbol{\theta}$ under the given measurement, when the precision is determined by the covariance matrix $\bm{\Sigma}$ ($\bm{\Sigma}_{\mu,\nu}=\mathrm{Cov}(\theta_{\mu},\theta_{\nu})$). The CFIM and covariance matrix both depend on the measurement applied. By optimization over all possible measurements, the CFIM itself is bounded by the QFIM $\boldsymbol{F}^q$ via the quantum Cram\'{e}r-Rao inequality (QCRI)
\begin{equation}
\bm{\Sigma}\geq{({\bm{F}^{c}})^{-1}}\geq{({\bm{F}^{q}})^{-1}},
\label{eq:CovFF}
\end{equation}
where the matrix elements of $\bm{F}^{q}$ is defined as
\begin{eqnarray}
\bm{F}^{q}_{\mu,\nu}&=&4\mathrm{Re}[\langle\partial_{\mu}\psi(\bm\theta)|\partial_{\nu}\psi(\bm\theta)\rangle \nonumber \\
&&\qquad-\langle\partial_{\mu}\psi(\bm\theta)|\psi(\bm\theta)\rangle \langle\psi(\bm\theta)|\partial_{\nu}\psi(\bm\theta)\rangle].
\label{eq:QFIM}
\end{eqnarray}
QFIM is solely determined by the parameterized output state and its dependence on the parameters.
Here, $\bm{F}^{c}$ and $\bm{F}^{q}$ are assumed invertible.

Following the standard methods, we extract a scalar  measure $\mathrm{tr}\boldsymbol{\Sigma}=\sum_\mu\delta^2\theta_\mu$ out of $\boldsymbol{\Sigma}$ to quantify the (inverse of the) estimation's precision. According to the QCRI, we have
\begin{equation}
\mathrm{tr}\boldsymbol{\Sigma}\geq\mathrm{tr}[(\boldsymbol{F}^c)^{-1}]\geq\mathrm{tr}[(\boldsymbol{F^q})^{-1}],\label{trQCRI}
\end{equation}
where the first equality can always be attained via the maximally likelihood estimation~\cite{Helstrom1967,Lu2020,Helstrom1976}. Thus, the \emph{precision} of an estimation scheme with a given measurement is measured by (inverse of) $\mathrm{tr}[(\boldsymbol{F^c})^{-1}]$. Furthermore, attainability of the second inequality in Eq.~\eqref{trQCRI} relies on the chosen measurement. It indicates that QFIM gives a lower bound of the {\textit {precision}} over all possible measurements. 
Based on above analysis, we measure the quality of a measurement method with the gap
\begin{equation}
\Delta=\mathrm{tr}[({\bm{F}^{c}})^{-1}]-\mathrm{tr}[({\bm{F}^{q}})^{-1}].
\label{eq:inequality}
\end{equation}
{A smaller $\Delta$ is thus indicates a more precise estimation, i.e., a better measurement method.}

\section{Two-parameter estimation with the STWS}

\subsection{The initial state preparation}

In quantum metrology, the quantum entanglement is the primary resource to improve measurement precision. It is well-known that the quantum entanglement in a two-mode NOON state can enhance the precision of single parameter estimation to the Heisenberg limit.
In this section, we will show that one can intialize our triple-well system to the high-entangeld \emph{three-mode NOON state}
\begin{equation}
\left|\psi\right\rangle_\mathfrak{i}=|\mathfrak{T}_0\rangle=\frac{1}{\sqrt{3}}\left(\left|e_1\right\rangle +\left|e_2\right\rangle +\left|e_3\right\rangle \right)
\label{eq:initialstate}
\end{equation}
in the adiabatic limit, with $|e_{1}\rangle=\left|N,0,0\right\rangle$, $|e_{2}\rangle=\left|0,N,0\right\rangle$, and $|e_{3}\rangle=\left|0,0,N\right\rangle$. We mention that $|\mathfrak{T}_0\rangle$ is optimal to estimate the relative phases between the three modes, when there is no  reference mode \cite{Goldberg2020}. 
\begin{figure}[tp]
{\includegraphics[width=75mm]{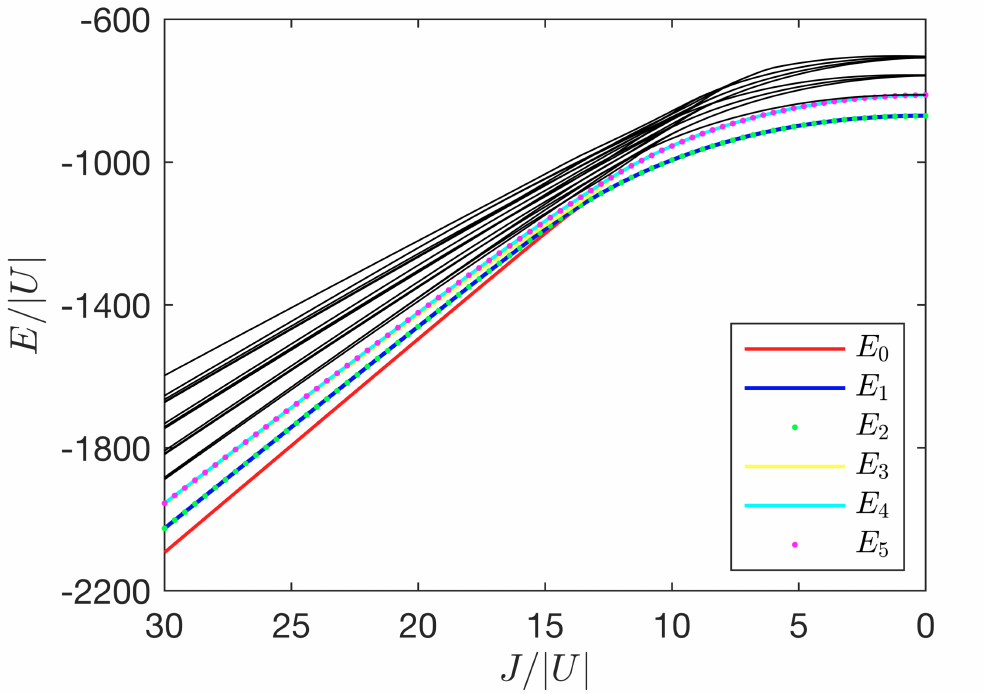}}
\caption{(Color online) Energy spectrum vs. $J$. Here we set $N=30$. Only the lowest 30 energy levels are given.
\label{fig:a}}
\end{figure}

Before introducing the preparation scheme, we briefly recall the eigenspectrum structure of the Hamiltonian Eq.~\eqref{eq:Hamiltonian}, which is shown in Fig.~\ref{fig:a}.
In the extreme case of $J=0$ and $U<0$, the energy level is completely arranged by the attractive interaction. It leads to a three-fold degenerate groundstate space $\mathcal{H}_N$ spanned by $|e_1\rangle$, $|e_2\rangle$, and $|e_3\rangle$, where all $N$ particles are located in a single trap.
When $J\neq 0$, the ground state space is separated into a single ground state and a two-dimensional excited space due to the tunneling term. The two-fold degeneration of the first and second excited states is induced by the chiral symmetry of the STWS.
In the strong coupling limit ($J/|NU|\gg1$), the tunneling term is dominant. The ground state and the two-fold degenerate excited states are well-separated by an energy gap which is proportional to the hopping strength $J$.

\begin{figure*}[tp]
\centering
\includegraphics[width=180mm]{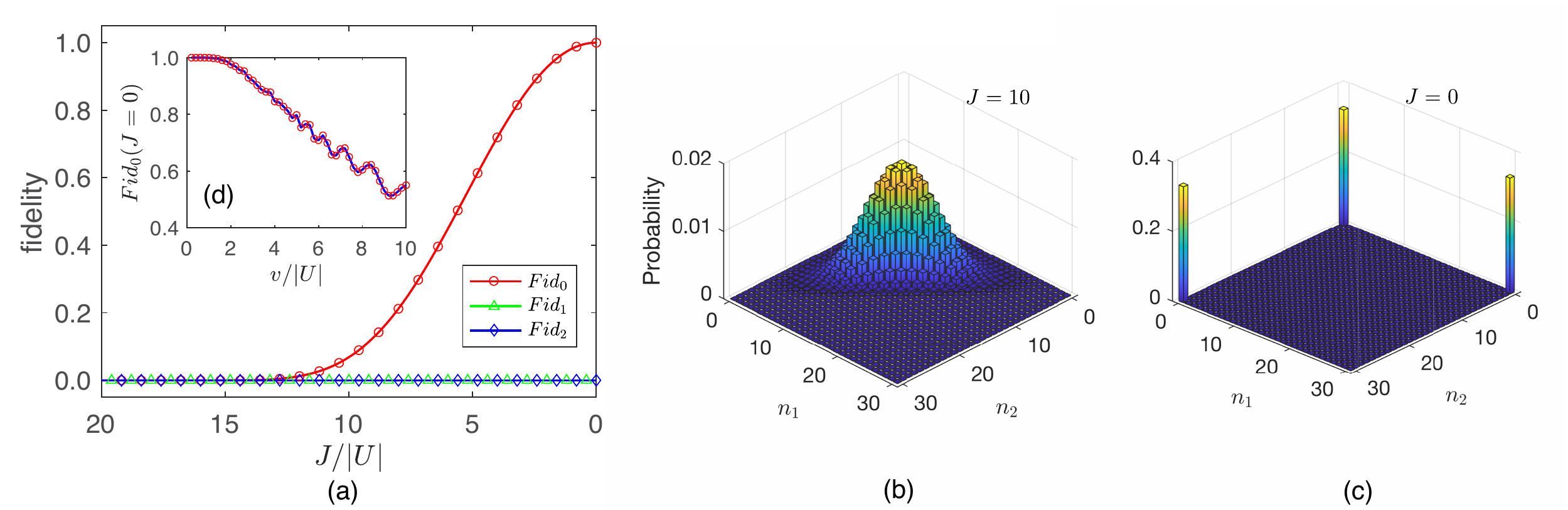}
\caption{(Color online)
State preparation. ($U=-0.5$ and $N=30$)
(a) The fidelity vs. $J$, where $Fid_a=|\langle\psi(t)|\mathfrak{T}_a\rangle|^2$, for $a=0,1,2$. $J=J_{0}-vt$ with $J_{0}=10$ and $v=0.2$.  $|\psi(t)\rangle$ is the evolving state from the ground state at $J=10$. Three-mode NOON state $|\mathfrak{T}_0\rangle$ is the target initial state. $\{|\mathfrak{T}_0\rangle, |\mathfrak{T}_1\rangle, |\mathfrak{T}_2\rangle\}$ are orthogonal basis of the there-dimensional ground state manifolds $\mathcal{H}_N$ at $J=0$.
(b-c) Distribution of quantum states in basis $|n_{1},n_{2},n_{3}\rangle$. (b) Ground state at $J=10$. (c) $|\psi(t)\rangle$ at $J=0$, with $v=0.2$. It is $|\mathfrak{T}_0\rangle$ approximately.  (d) The inset shows the speed dependence of the state preparation via the final state fidelity $Fid_0$ at $J=0$ vs. the speed $v$.
\label{fig:b}}
\end{figure*}

Next, we introduce the preparation of the target initial state.
Firstly, we prepare STWS into the ground state $|\psi_{J_0}\rangle_g$ in the strong coupling limit$(J_0/|NU|\gg1)$ at $t=0$. It is manageable due to the large energy gap. Then we slowly decrease the coupling strength $J$ to zero with $J=J_0-vt$. The evolving state is thus given by
\begin{equation}
|\psi(t_1)\rangle=\mathcal{T}\mathrm{exp}[-{\rm i}\int_0^{t_1}\hat{H}(J)\mathrm{d}t]|\psi_{J_0}\rangle_g.
\end{equation}
We adjust the decreasing rate $v$ according to the adiabatic fidelity \cite{Liu2007}.
The ground state will evolve into the three-mode NOON state $|\mathfrak{T}_0\rangle$ at $J=0$, if the decreasing rate is lower enough to meet the adiabatic condition. One can see the particle distribution changes from Fig.~\ref{fig:b}(b) to Fig.~\ref{fig:b}(c), equally distributed in $|e_1\rangle$, $|e_2\rangle$, and $|e_3\rangle$. Furthermore, the vanishing of the relative phases is verified numerically.

Intuitively speaking, this adiabatic process naturally presents a state in the three-dimensional ground state space $\mathcal{H}_N$ at $J=0$ in the adiabatic limit. The subtle point is that the state we prepared is $|\mathfrak{T}_0\rangle$ exactly. To show this result more rigorously, we introduce three basis states of the ground state space $\mathcal{H}_N$, which reads $|\mathfrak{T}_0\rangle$, and  $|\mathfrak{T}_{1(2)}\rangle=(|e_1\rangle+e^{\pm {\rm i}2\pi/3}|e_2\rangle+e^{\mp {\rm i}2\pi/3}|e_3\rangle)/\sqrt{3}$, respectively.
Then we calculate the fidelity $Fid_a=|\langle\psi(t)|\mathfrak{T}_a\rangle|^2$ for the whole adiabatic process, with $a=0, 1, 2$, respectively. The result is plotted as Fig.~\ref{fig:b}(a).
It shows $Fid_0$ increases with decreasing $J$. Finally, we prepare the target state $|\mathfrak{T}_0\rangle$ at $J=0$ almost certainly, as indicated by $Fid_0\approx 1$. Meanwhile, fidelities of the other two states, $Fid_1$ and $Fid_2$, vanish not only for $J=0$ but all $J$. It indicates that $|\mathfrak{T}_1\rangle$ and $|\mathfrak{T}_2\rangle$, hence, all of states in $\mathcal{H}_N$ other than $|\mathfrak{T}_0\rangle$, are excluded in the whole adiabatic evolution.
 It is strong evidence that the state $|\psi\rangle_\mathfrak{i}$ we prepared is precisely the target state $|\mathfrak{T}_0\rangle$ in the adiabatic limit.

\subsection{Parameterization and the QFI}

In the last section, the three-mode NOON state $|\mathfrak{T}_0\rangle$ is prepared as the initial state of STWS.  While keeping $J=0$, we put the STWS in state $|\mathfrak{T}_0\rangle$ into an external field. Denote the shifted energy of mode $\hat{a}_i$ (ground state energy of the $i$-th trap) in this field as $\mathcal{E}_i$. Then the mode $\hat{a}_i$ will evolve as $\hat{a}_ie^{i\phi_i}$ with $\phi_i=\mathcal{E}_i t$, after time $t$. The state $\left|\psi\right\rangle_\mathfrak{i}$ is thus parameterized to the output state as
\begin{eqnarray}
\left|\psi(\bm{\theta})\right\rangle
&=&\frac{1}{\sqrt{3}}({\rm e}^{-\mathrm{i}N\theta_{1}}\left|e_1\right\rangle +{\rm e}^{-\mathrm{i}N\theta_{2}}\left|e_2\right\rangle +\left|e_3\right\rangle ).
\label{eq:psiout}
\end{eqnarray}
where $\bm{\theta}=(\theta_{1},\theta_{2})$, with $\theta_i=\phi_i-\phi_3$, is the vector parameter to be estimated. We mention that the on-site interaction is negligible for it only contributes a global phase to $|\psi(\boldsymbol{\theta})\rangle$.

{Substituting Eq.~\eqref{eq:psiout} into Eq.~\eqref{eq:QFIM}, we can calculate the four matrix elements of the QFIM on $\bm{\theta}$ directly.  The corresponding QFIM is}
\begin{equation}
\bm{F}^{q}=\frac{4N^{2}}{9}\left(\begin{array}{cc}
2 & -1\\
-1 & 2
\end{array}\right).
\end{equation}
It shows that the QFIM only depends on the total particle number $N$, independent of $\bm{\theta}$.
Furthermore, based on Eq.~\eqref{trQCRI}, we have 
\begin{equation}
\langle\Delta^{2}\bm{\theta}\rangle\geq \mathrm{tr}{[(\bm{F}^{q})^{-1}]}=\frac{3}{N^2},
\label{eq:CovFq}
\end{equation}
where $\langle\Delta^{2}{\bm{\theta}}\rangle\equiv\langle\Delta^{2}{\theta_{1}}\rangle+\langle\Delta^{2}{\theta_{2}}\rangle$ is the total variance of $\theta_{1}$ and $\theta_{2}$. 
Eq.~\eqref{eq:CovFq} indicates that
the upper bound precision of estimating $\bm{\theta}$ can approach the Heisenberg scaling.

\subsection{Projective measurement}

We have discussed the theoretical limit of the sensitivity via the QFIM. However,
the accessible sensitivity highly depends on the measurement scheme.
In this section, we focus on particle number measurement and study its precision with the CFIM.

In STWS, it is convenient to measure the particle number in each well on the final state. This measurement is depicted by a set of projection operators
\begin{equation}
\{\hat{\Pi}_{\bm{n}}\}=\{|n_{1},n_{2},n_{3}\rangle\langle n_{1},n_{2},n_{3}|\},
\label{eq:projector}
\end{equation}
with $\bm{n}=(n_{1},n_{2},n_{3})$.
However, the phases will be eliminated if we directly perform this measurement on the parameterized state $|\psi(\boldsymbol{\theta})\rangle$. In this way, the parameter $\bm{\theta}$ cannot be inferred.
Hence, as the pretreatment of the measurement, we rotate the output state $|\psi(\boldsymbol{\theta})\rangle$ with
\begin{equation}
\hat{\mathcal{U}}_{R}(\tau)=\mathrm{exp}[-\mathrm{i}\hat{H}_{R} \tau/J],
\label{eq:UR}
\end{equation}
 where $\tau=Jt$ is the rescaled rotation time and
\begin{equation}
\hat{H}_{R}\approx-J\sum_{i=1}^{3}(\hat{a}_{i}^{\dagger}\hat{a}_{j}+h.c.),
\label{eq:HR}
\end{equation}
with $j=(i+1)$mod$3+1$. The tunneling strength $J$ is fixed throughout the rotation.
This Hamiltonian is valid when $J\gg |U|N$.
It can be realized by both 1) increasing $J$ via lowering the energy barrier between two traps and 2) tuning $U\approx 0$ via the Feshbach resonance simultaneously.
Based on Eqs.~\eqref{eq:psif}, \eqref{eq:psiout}, and \eqref{eq:UR}, we have the rotated final state as (see Appendix A)
\begin{equation}
|\psi\rangle_{\mathfrak{f}} = C\eta(\tau)^{N} \sum_{n_{1},n_{2}}\sum_{i=1}^{3}{\rm e}^{-\mathrm{i} N\theta_{i}}
\xi(\tau)^{n_{i}}|n_{1},n_{2},n_{3}\rangle
\label{eq:finalstate}
\end{equation}
with $C=(\frac{N!}{3^{2N+1}n_{1}!n_{2}!n_{3}!})^{1/2}$,  $\xi(\tau)=({\rm e}^{3\mathrm{i}\tau}+2)/({\rm e}^{3\rm{i}\tau}-1)$, $\eta(\tau)={\rm e}^{2\mathrm{i}\tau}-{\rm e}^{-\mathrm{i}\tau}$, $\theta_3=0$.

Applying the particle number measurement $\hat{\Pi}_{\bm{n}}$ on the final state $|\psi\rangle_\mathfrak{f}$, we have the probability of acquiring result $|n_1,n_2,n_3\rangle$ as
\begin{equation}
P_{\bm n}=C^2\left[2\sin(\frac{3\tau}{2})\right]^{2N}\left|\sum_{i}{\rm e}^{-{\rm i}N\theta_i}\xi(\tau)^{n_i}\right|^2 .
\label{eq:Proba}
\end{equation}
The CFIM of probability $P_{\bm n}$ can be acquired directly via the definition Eq.~\eqref{eq:Fc}. We denote it as $\boldsymbol{F}^c(\boldsymbol{\theta},\tau)$, for it depends on both the estimated vector $\bm{\theta}$ and $\tau$. 
\subsection{The optimal measurement precision}

In this section, we will analyze the measurement precision with the gap $\Delta$ defined in Eq.~\eqref{eq:inequality}. The optimal precision can be given by  optimizing both the encoded parameters $\boldsymbol{\theta}$ and rotation time $\tau$.

\begin{figure}[t]
\centering
\includegraphics[width=80mm]{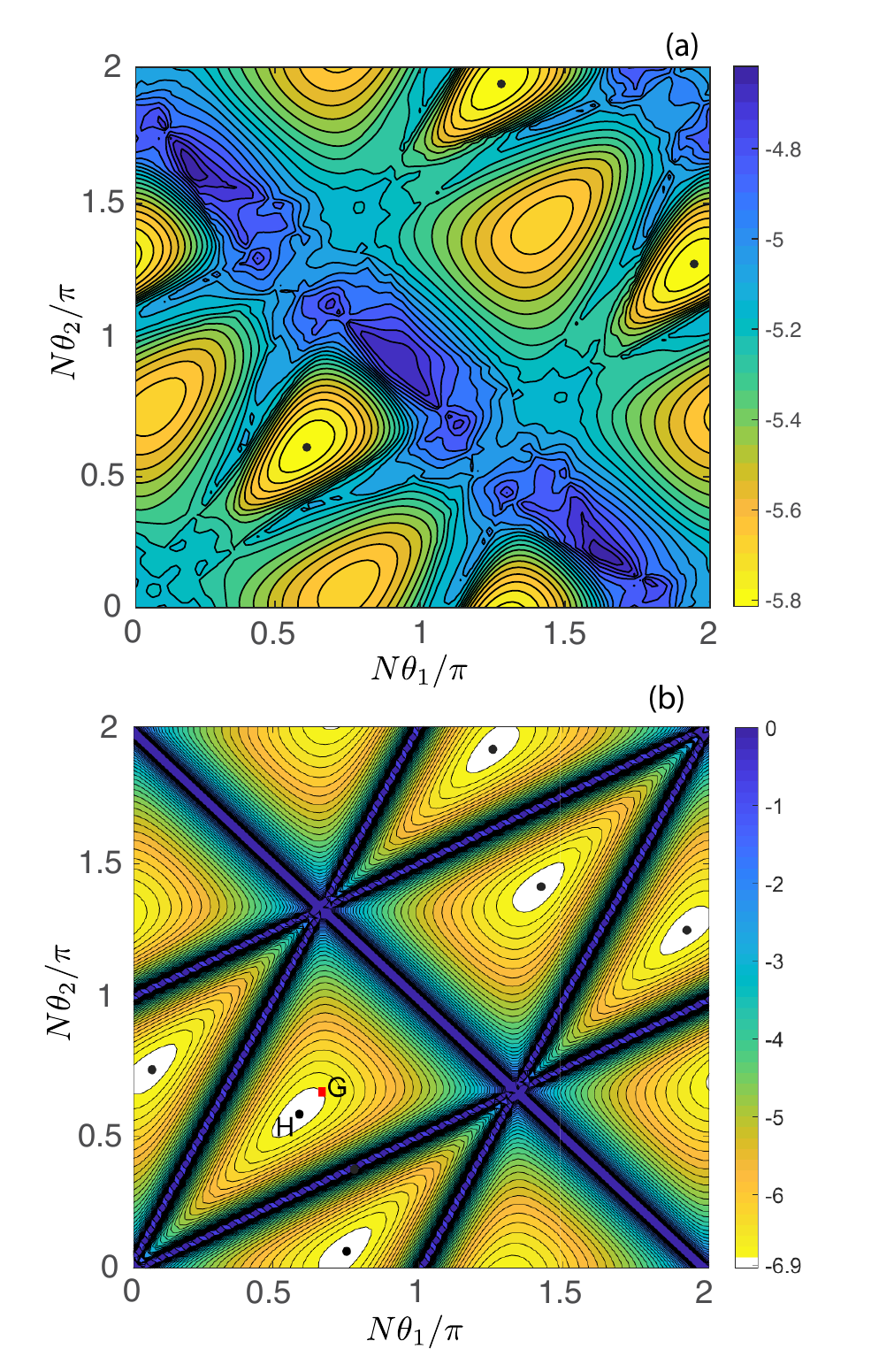}
\caption{(Color online) $\ln(\Delta)$ vs. $(\theta_1,\theta_2)$ with a given $\tau$.
(a) $\tau=0.2\pi$. (b)$\tau=\tau_O=2\pi/9$. The maximum precision (minimum of $\ln(\Delta)$) for a given $\tau$ is marked as the black dots "$\bullet$". $\boldsymbol{\theta}_G=(2\pi/3,2\pi/3)$. Here, we set $N=30$.}
\label{fig:cc}
\end{figure}
\begin{figure}[t]
\includegraphics[width=80mm]{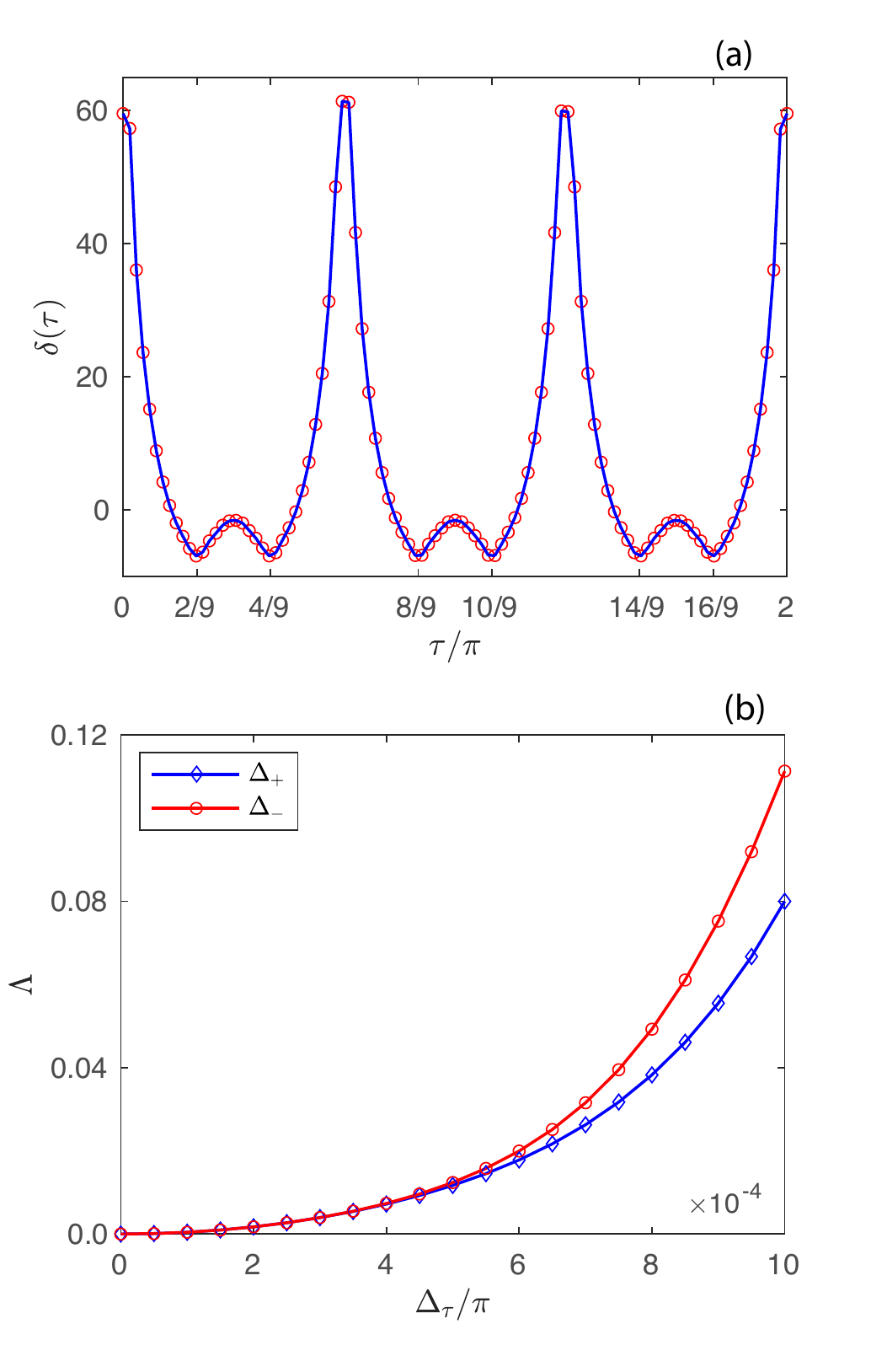}
\caption{(Color online) (a) $\delta(\tau)$ vs. $\tau$. (b) $\Lambda$ vs. $\Delta_\tau$. $\Lambda=\delta(\tau)-\delta(\tau_O)$, ''$\Delta\pm$'' denote the line for $\tau =\tau_O\pm\Delta_\tau$, with $\tau_O=2\pi/9$.
$\Lambda$ decreases to zero from above as $\Delta_\tau$ approaches $0$. Here, we set $N=30$. }
\label{fig:d}
\end{figure}

\begin{figure}[tb]
{\includegraphics[width=78mm]{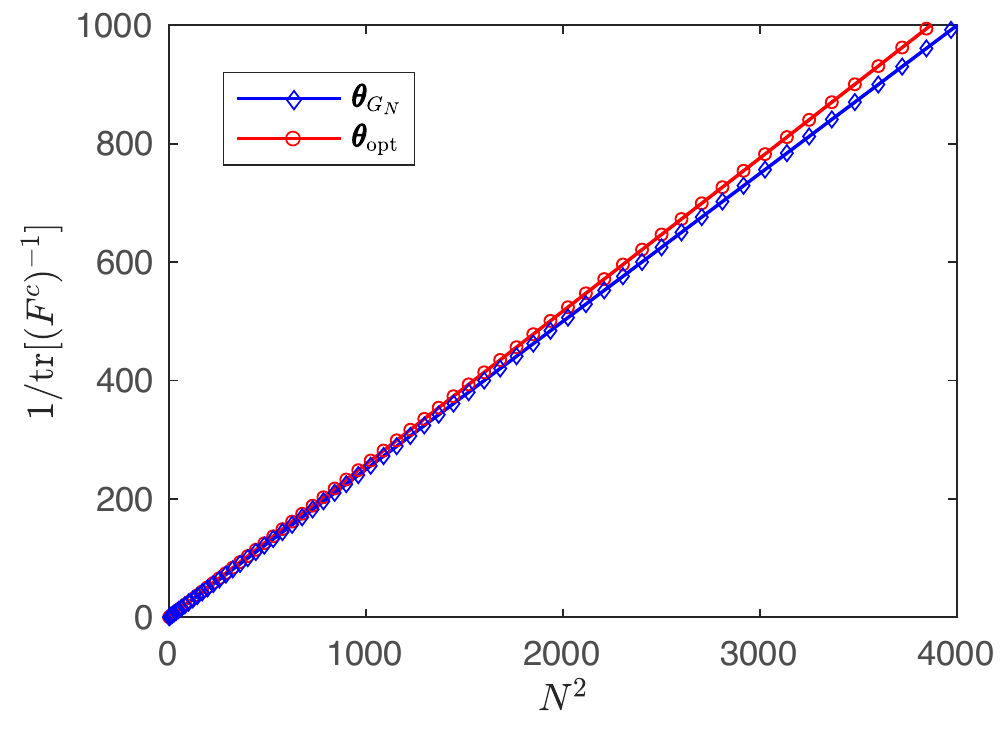}}
\caption{(Color online) The precision ($1/\mathrm{tr}[(\bm{F}^{c})^{-1}]$) vs. $N^2$.  Here, we set $\tau=\tau_O=2\pi/9$. The red line denotes the numerical optimal precision. The blue line is analytical result at $\boldsymbol{\theta}_{G_N}$, which reads $\mathrm{tr}[(\bm{F}^{c})^{-1}]= {4}/{N^2}$. }
\label{fig:fcN}
\end{figure}

Firstly, we discuss the dependence of $\Delta$ on $\boldsymbol{\theta}$ with a given rotation time $\tau$ numerically.
The results are shown in Fig.~\ref{fig:cc}, where only one period is given.
We observe that $\Delta$ highly depends on $\theta_{1}$ and $\theta_{2}$. For a given $\tau$, the maximal precision can only be achieved in several points "$\bullet$". Luckily, with enough prior information provided, we can shift the estimand to the vicinity of these points to approach the maximal precision.

Secondly, by comparing Fig.~\ref{fig:cc}(a) and Fig.~\ref{fig:cc}(b), we find that the maximal precision over parameters $\boldsymbol{\theta}$ highly depends on the rotation time $\tau$. 
Specifically, we define
\begin{equation}
\delta(\tau)=\min_{\{\theta_{1},\theta_{2}\}}^{}[\ln(\Delta)]
\label{eq:deltatau}
\end{equation}
to study the effect of rotation time $\tau$ on the optimal precision over $\boldsymbol{\theta}$.
As shown by Fig.~\ref{fig:d}(a), $\delta(\tau)$ varies periodically with $\tau$. There are three short periods with duration $2\pi/3$ in a long period with duration $2\pi$.
More importantly, $\delta(\tau)$ takes the minimum value at  points
\begin{equation}
\tau\approx\frac{(2k+1)\pi}{3}\pm\frac{\pi}{9},
\label{eq:Rtime}
\end{equation}
with $k\in\mathbb{N}$. We further show its validity numerically in Fig.~\ref{fig:d}(b), which indicates that $\tau_O=2\pi/9$ is one of the optimal rotation times.

Now, one can acquire the optimal precision of our scheme by choosing the optimal phases $(\theta_{1},\theta_{2})$ at an optimal time $\tau$ given in Eq.~(\ref{eq:Rtime}).
It can be done numerically. An example with $\tau=\tau_O=2\pi/9$ is given in Fig.~\ref{fig:cc}(b), where the point "H" is one of the optimal sets of phases. 

To evaluate the quality of the optimal precision, we study the scaling of CFIM with particle number $N$ at $\tau=\tau_O={2\pi}/{9}$ both numerically and analytically.
The numerical result is shown as the red line in Fig.~\ref{fig:fcN}. By searching the minimum of $\mathrm{tr}[(\boldsymbol{F}^c)^{-1}]$  over the phase parameter $\boldsymbol{\theta}$ at $\tau_O={2\pi}/{9}$, we find the optimal precision
satisfies the following linear relationship
\begin{equation}
\min_{\{\theta_1\theta_2\},\tau_O}\mathrm{tr}[(\bm{F}^{c})^{-1}] \approx 3.87\times  \frac{1}{N^2}.
\label{eq:trFC}
\end{equation}
It indicates a Heisenberg scaling precision.
To show it more concretely, we further give a lower bound of the optimal precision analytically. The precision for the optimal phase point at $\tau_O$ is challenging to be given analytically. Hence, we calculate the precision for point "G" instead, which is located near the optimal point "H" The phase parameter $\boldsymbol{\theta}_{G_N}$ of the point "G" is given by $N\theta_1=N\theta_2=2(N+1)\pi/3$, with $N$ denoting the particle number. The CFIM of the state “G" is (see Appendix~B)
\begin{eqnarray}
\bm{F}^{c}(\boldsymbol{\theta}_{G_N},\frac{2\pi}{9})=\frac{N^2}{3}\left(\begin{array}{cc}
 2 & -1\\
 -1 & 2
\end{array}\right).
 \label{eq:theta12}
\end{eqnarray}
And the corresponding precision is
\begin{equation}
\mathrm{tr}[(\boldsymbol{F}^c)^{-1}]=\frac{4}{N^2}.
\label{eq:tra}
\end{equation}
The result is shown as the blue line in Fig.~\ref{fig:fcN} in comparison with the optimal precision given by Eq.~(\ref{eq:trFC}).  It shows that scaling of the two precisions are very close.

We mention that Eq.~(\ref{eq:tra}) is valid for all particle numbers $N$. It indicates that, with the proposed measurement scheme, the optimal measurement precision of $\boldsymbol{\theta}$ can always show a better Heisenberg scaling than Eq.~(\ref{eq:tra}). 
Furthermore, as indicated by Fig.~\ref{fig:cc}(b) and Fig.~\ref{fig:fcN}, the precision is robust around the optimal parameters, e.g., $|\psi(\boldsymbol{\theta}_G)\rangle$ with parameters $\boldsymbol{\theta}_G$ still have relatively high precision with a Heisenberg scaling. It significantly reduces the demands for practical studies, which relieve the working parameters' constrain from a point "H" to, e.g., the white zoom in Fig.~\ref{fig:cc}(b). With a relatively larger acceptance zoom of the parameter shifts,  the demands for the prior information about the estimand $\boldsymbol{\theta}$ are also highly reduced.

\begin{figure}[tp]
{\includegraphics[width=78mm]{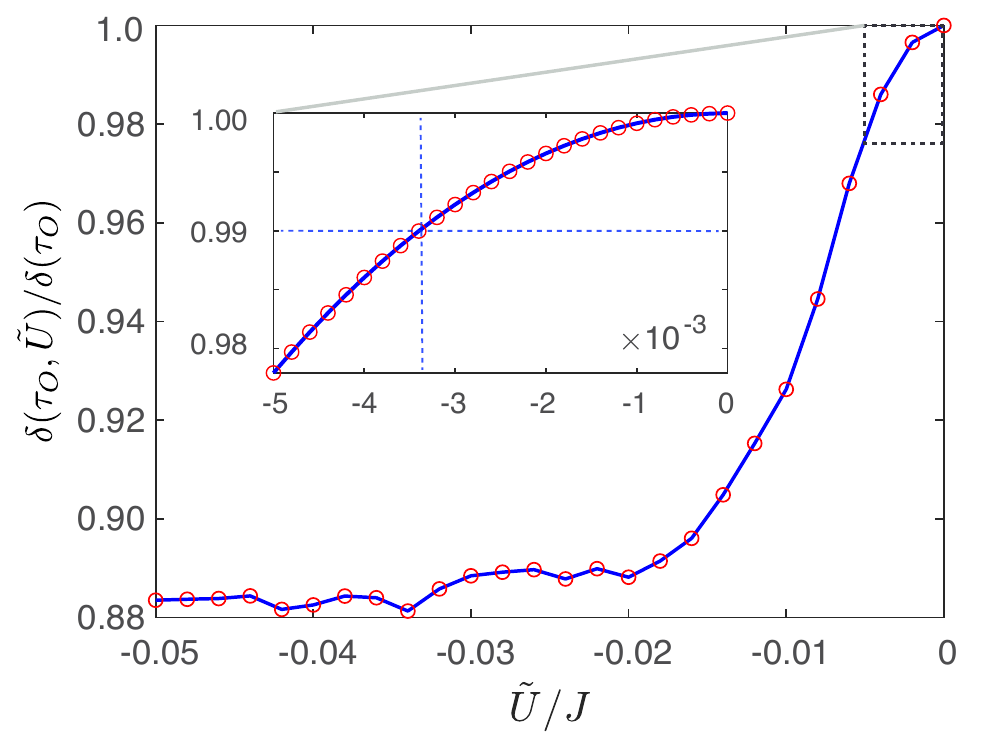}}
\caption{(Color online)
The precision $\delta(\tau_O,\tilde{U})$ vs. $\tilde{U}$, with $\tau_O=2\pi/9$.
$\delta(\tau_O,\tilde{U})$ is the optimal precision acquired via the imperfect rotation $\hat{\mathcal{U}}'_{R}(\tau_O)$.
$\delta(\tau_O)=\delta(\tau_O,0)$.
Here, we take $J=10$, $N=30$.
 \label{fig:U}}
\end{figure}

We have discussed the optimal precision under the projection measurement. However, the result is acquired under the approximation Eq.~\eqref{eq:HR}. If the on-site interaction between atoms  cannot be tuned to zero precisely in the rotation operation, the total Hamiltonian reads   
\begin{equation}
\hat{H}_{R}'=-J\sum_{i=1}^{3}(\hat{a}_{i}^{\dagger}\hat{a}_{j}+h.c.)+\tilde{U}\sum_{i=1}^{3}\hat{n}_{i}(\hat{n}_{i}-1),
\label{eq:HU}
\end{equation}
where $\tilde{U}$ denotes a small residual interaction induced by the imperfect control of the Feshbach resonance.
To discuss the effect of residual interaction on the optimal precision, we define $\delta(\tau_O,\tilde{U})$ as the generalization of $\delta(\tau_O)$ (Eq.~(\ref{eq:deltatau})) by substituting $\hat{\mathcal{U}}_R(\tau_O)$ to the imperfect rotation
\begin{equation}
\hat{\mathcal{U}'}_{R}(\tau_O)=\exp[-{\rm i}\hat{H}_R'\tau_O/J].
\end{equation}
$\delta(\tau_O,\tilde{U})$ give us the optimal precision over $\boldsymbol{\theta}$ acquired under the imperfect rotation.
We plot the ratio $\delta(\tau_O,\tilde{U})$ as a function of $\tilde{U}$ in Fig.~\ref{fig:U}. Although $\delta(\tau_O,\tilde{U})$ decreases with increasing $\tilde{U}$, the precision $\delta(\tau_O,\tilde{U})$ is roughly on $99\%$ of $\delta(\tau_O)$ when $|N\tilde{U}/J|=0.1$. Even when $|N\tilde{U}/J|\approx 1$, i.e., the residual interactions and tunneling energy are on the same level, there still exists a platform around $\delta(\tau_O,\tilde{U})=0.89 \delta(\tau_O)$.  
Thus, imperfect rotation with small residue interaction $\tilde{U}$ has acceptable influence on the optimal precision.

\section{Conclusion and discussion}

In this work, we have proposed a scheme for two-parameter estimation  via a Bose-Einstein condensate confined in a symmetric triple-well potential.
The three-mode NOON state has been prepared adiabatically as the initial state.
The two parameters to be estimated are the two phase differences between the wells, which are encoded into the initial state via the external fields. We perform the particle number measurement in each well to read the parameterized state. Moreover, a rotation operation is adopted on the output state before the measurement.
We optimize both the parameters and rotation time to maximize the estimation precision.  As a result, we have approached the Heisenberg scaling precision on simultaneous estimating two parameters under the optimal measurement.

We mention that our scheme is discussed in the ideal scenario in this article. To study it more rigorously, one should build an open quantum system model and introduce noise analysis based on practical experiments. We will advance this research in further studies. 
We expect to realize the high precision estimation of the two-dimensional fields, such as the magnetic field and gravity field, via ongoing research in this triple-well system.

\section{Acknowledgments}

Fei Yao thanks Peng Wang for the helpful discussion.
This work was supported by the National Natural Science Foundation of China (NSFC) (No.~12088101, Grant No.~11725417, and Grant No.~U1930403) and Science Challenge Project (Grant No. TZ2018005).

\begin{widetext}

\appendix

\section{The derivation of the rotated final state}

The parameterized output state is
\begin{equation}
\left|\psi(\bm{\theta})\right\rangle =\frac{1}{\sqrt{3}}\left(\exp\left[-\mathrm{i}N\theta_{1}\right]\left|N,0,0\right\rangle +\exp\left[-\mathrm{i}N\theta_{2}\right]\left|0,N,0\right\rangle +\left|0,0,N\right\rangle \right).
\end{equation}
The output state will be rotated by the operation $\hat{\mathcal{U}}_{R}=\mathrm{exp}[-\mathrm{i}\hat{H}_{R} t]$.
The Hamiltonian $\hat{H}_{R}=- J\sum_{i,j=1,i\neq j}^{3}\hat{a}_{i}^{\dagger}\hat{a}_{j}$ can be  diagonalized in the basis
\begin{equation}
\left(\begin{array}{c}
\alpha_{1}\\
\alpha_{2}\\
\alpha_{3}
\end{array}\right)=\frac{1}{\sqrt{3}}\left(\begin{array}{ccc}
1 & 1 & 1\\
1 & \mathrm{e^{\mathrm{i}2\pi/3}} & \mathrm{e^{-\mathrm{i}2\pi/3}}\\
\mathrm{1} & \mathrm{e^{-\mathrm{i}2\pi/3}} & \mathrm{e^{\mathrm{i}2\pi/3}}
\end{array}\right)\left(\begin{array}{c}
a_{1}\\
a_{2}\\
a_{3}
\end{array}\right)
\label{eq:trans}
\end{equation}
to give
\begin{equation}
\hat{H}_{R}=J\left(-2\alpha_{1}^{\dagger}\alpha_{1}+\alpha_{2}^{\dagger}\alpha_{2}+\alpha_{3}^{\dagger}\alpha_{3}\right).
\end{equation}
The output state $\left|\psi(\bm{\theta})\right\rangle$ can be transformed into the eigenbasis vector of $\hat{H}_{R}$
via Eq.~\eqref{eq:trans}.
The transformation of $\left|\psi(\bm{\theta})\right\rangle$ is thus given as below,
\begin{eqnarray}
\left|\psi\right\rangle _{f} & = & \hat{\mathcal{U}}_{R} \left|\psi(\bm{\theta})\right\rangle\nonumber\\
& = & \exp\left[-\mathrm{i}Jt\left(-2\alpha_{1}^{\dagger}\alpha_{1}+\alpha_{2}^{\dagger}\alpha_{2}+\alpha_{3}^{\dagger}\alpha_{3}\right)\right]\left|\psi(\bm{\theta})\right\rangle \nonumber\\
 & = & \exp\left[-\mathrm{i}Jt\left(-2\hat{n}_{\alpha1}+\hat{n}_{\alpha2}+\hat{n}_{\alpha3}\right)\right]\frac{1}{\sqrt{3^{N+1}N!}}\{\exp\left[-\mathrm{i}N\theta_{1}\right]\left(\alpha_{1}^{\dagger}+\alpha_{2}^{\dagger}+\alpha_{3}^{\dagger}\right)^{N}\left|0,0,0\right\rangle \nonumber \\
 &  & +\exp\left[-\mathrm{i}N\theta_{2}\right]\left(\alpha_{1}^{\dagger}+\mathrm{e^{\mathrm{i}2\pi/3}}\alpha_{2}^{\dagger}+\mathrm{e^{-\mathrm{i}2\pi/3}}\alpha_{3}^{\dagger}\right)^{N}\left|0,0,0\right\rangle +\left(\alpha_{1}^{\dagger}+\mathrm{e^{-\mathrm{i}2\pi/3}}\alpha_{2}^{\dagger}+\mathrm{e^{\mathrm{i}2\pi/3}}\alpha_{3}^{\dagger}\right)^{N}\left|0,0,0\right\rangle \} \nonumber\\
 & = & \frac{1}{\sqrt{3^{N+1}N!}}\{\exp\left[-\mathrm{i}N\theta_{1}\right]\left(\mathrm{e}^{2\mathrm{i}Jt}\alpha_{1}^{\dagger}+\mathrm{e}^{-\mathrm{i}Jt}\alpha_{2}^{\dagger}+\mathrm{e}^{-\mathrm{i}Jt}\alpha_{3}^{\dagger}\right)^{N}\left|0,0,0\right\rangle \nonumber \\
 &  & +\exp\left[-\mathrm{i}N\theta_{2}\right]\left(\mathrm{e}^{2\mathrm{i}Jt}\alpha_{1}^{\dagger}+\mathrm{e^{\mathrm{i}2\pi/3}}\mathrm{e}^{-\mathrm{i}Jt}\alpha_{2}^{\dagger}+\mathrm{e^{-\mathrm{i}2\pi/3}}\mathrm{e}^{-\mathrm{i}Jt}\alpha_{3}^{\dagger}\right)^{N}\left|0,0,0\right\rangle \nonumber \\
 &  & +\left(\mathrm{e}^{2\mathrm{i}Jt}\alpha_{1}^{\dagger}+\mathrm{e^{-\mathrm{i}2\pi/3}}\mathrm{e}^{-\mathrm{i}Jt}\alpha_{2}^{\dagger}+\mathrm{e^{\mathrm{i}2\pi/3}}\mathrm{e}^{-\mathrm{i}Jt}\alpha_{3}^{\dagger}\right)^{N}\left|0,0,0\right\rangle \}\nonumber\\
 & = & \frac{1}{\sqrt{3^{2N+1}N!}}\{\exp\left[-\mathrm{i}N\theta_{1}\right]\left[\left(\mathrm{e}^{2\mathrm{i}Jt}+2\mathrm{e}^{-\mathrm{i}Jt}\right)a_{1}^{\dagger}+\left(\mathrm{e}^{2\mathrm{i}Jt}-\mathrm{e}^{-\mathrm{i}Jt}\right)\left(a_{2}^{\dagger}+a_{3}^{\dagger}\right)\right]^{N}\left|0,0,0\right\rangle \nonumber \\
 &  & +\exp\left[-\mathrm{i}N\theta_{2}\right]\left[\left(\mathrm{e}^{2\mathrm{i}Jt}+2\mathrm{e}^{-\mathrm{i}Jt}\right)a_{2}^{\dagger}+\left(\mathrm{e}^{2\mathrm{i}Jt}-\mathrm{e}^{-\mathrm{i}Jt}\right)\left(a_{1}^{\dagger}+a_{3}^{\dagger}\right)\right]^{N}\left|0,0,0\right\rangle \nonumber \\
 &  & +\left[\left(\mathrm{e}^{2\mathrm{i}Jt}+2\mathrm{e}^{-\mathrm{i}Jt}\right)a_{3}^{\dagger}+\left(\mathrm{e}^{2\mathrm{i}Jt}-\mathrm{e}^{-\mathrm{i}Jt}\right)\left(a_{1}^{\dagger}+a_{2}^{\dagger}\right)\right]^{N}\left|0,0,0\right\rangle \}\nonumber\\
 & = & \sum_{n_{1},n_{2}}\sum_{i=1}^{3}\sqrt{\frac{N!}{3^{2N+1}n_{1}!n_{2}!n_{3}!}}\exp\left[-\mathrm{i}N\theta_{i}\right]\left(\mathrm{e}^{2\mathrm{i}Jt}+2\mathrm{e}^{-\mathrm{i}Jt}\right)^{n_{i}}\left(\mathrm{e}^{2\mathrm{i}Jt}-\mathrm{e}^{-\mathrm{i}Jt}\right)^{N-n_{i}}\left|n_{1},n_{2},n_{3}\right\rangle,
\end{eqnarray}
with $\theta_{3}=0$, $n_{1}+n_{2}+n_{3}=N$. It is equivalent to Eq.~(\ref{eq:finalstate}).

\section{Scalling of the CFIM entries}

In this section, we will show that: For a final state $|\psi\rangle_\mathfrak{f}$ with $\tau=2\pi/9$, there exist parameters $\boldsymbol{\theta}_{G_N}$ with $N\theta_1=N\theta_2=(N+1)2\pi/3$, such that the QFI entries
\begin{equation}
F_{11}^{c}=F_{22}^{c}=\frac{2N^{2}}{3},F_{12}^{c}=-\frac{N^{2}}{3}.\label{Ent}
\end{equation}
We begin with the cases where the total particle number $N=3k$, $k\in\mathbb{N}^+$.
Set $N\theta_{1}=N\theta_{2}=2\pi/3$, we have the probability of acquiring result $|n_1,n_2,n_3\rangle$ as
\begin{align}
P_{\boldsymbol{n}}
= & \frac{N!}{3^{N+1}n_{1}!n_{2}!n_{3}!}\left\{ 3+2\cos[\frac{2\pi}{3}(n_{2}-n_{3}+1)]+2\cos[\frac{2\pi}{3}(n_{1}-n_{3}+1)]+2\cos[\frac{2\pi}{3}(n_{1}-n_{2})]\right\},
\end{align}
which is equivalent to Eq.~(\ref{eq:Proba}). And the derivatives read
\begin{align}
\partial_{1}P_{\boldsymbol{n}}
= & \frac{N!N}{3^{N+1}n_{1}!n_{2}!n_{3}!}\left\{ -2\sin[\frac{2\pi}{3}(n_{1}-n_{3}+1)]-2\sin[\frac{2\pi}{3}(n_{1}-n_{2})]\right\}, \\
\partial_{2}P_{\boldsymbol{n}}
= & \frac{N!N}{3^{N+1}n_{1}!n_{2}!n_{3}!}\left\{ -2\sin[\frac{2\pi}{3}(n_{2}-n_{3}+1)]+2\sin[\frac{2\pi}{3}(n_{1}-n_{2})]\right\}.
\end{align}
For simplicity, we further reformulate the CFIM entries as
\begin{equation}
F_{\mu\nu}^{c}=N^{2}\sum_{n_{1},n_{2}}\frac{N!}{3^{N+1}n_{1}!n_{2}!n_{3}!}f_{\mu\nu}(\boldsymbol{n}),\label{eq:FElem}
\end{equation}
with
\begin{equation}
f_{\mu\nu}(\boldsymbol{n})=\frac{3^{N+1}n_{1}!n_{2}!n_{3}!}{N!N^{2}}\frac{\partial_{\mu}P_{\boldsymbol{n}}\partial_{\nu}P_{\boldsymbol{n}}}{P_{\boldsymbol{n}}}.
\end{equation}
The value of $f_{\mu\nu}(\boldsymbol{n})$ only depend on $\boldsymbol{n}\mathrm{mod}3$. We list them in Tab.~\ref{Tab1}.
\begin{table}[h]
\centering
\caption{\label{Tab1} Classification of $f_{\mu\nu}(\boldsymbol{n})$ with $\boldsymbol{n}\mod3$.}
\renewcommand{\arraystretch}{1.1}
\begin{tabular}{ccccccccccc}
\hline
\hline
\multirow{2}{*}{$\boldsymbol{n}\mod3$} & \{0,0,0\} & \{1,1,1\} &\multicolumn{6}{c}{\{0,1,2\}}\\
\cmidrule(lr){4-9}
 & (0,0,0)  & (1,1,1) & (0,1,2) & (0,2,1)  & (1,0,2)  & (1,2,0)  & (2,0,1)  & (2,1,0)\\
\hline
$f_{11}(\boldsymbol{n})$  & 1 & 1 & 4  & 1  & 1  & 4  & 4  & 1\\
$f_{12}(\boldsymbol{n})$  & 1 & 1 & -2 & -2 & -2 & -2 & -2 & -2\\
$f_{22}(\boldsymbol{n})$  & 1 & 1 & 1  & 4  & 4  & 1  & 1  & 4\\
\hline
\hline
\end{tabular}
\end{table}
We mention that the coefficient of $f_{\mu\nu}{(\boldsymbol{n})}$ in Eq.~(\ref{eq:FElem}) is invariant under the arbitrary permutation of $\{n_1,n_2,n_3\}$.
Hence,  $f_{\mu\nu}(\boldsymbol{n})$ can be further divided into three types with $\{n_1, n_2, n_3\} \mod 3=\{0, 0, 0\}$, $\{1, 1, 1\}$, and $\{0, 1, 2\}$, respectively.  And $F_{\mu\nu}^{c}$ is invariant under the substitution
\begin{align}
f_{11}(\boldsymbol{n}),f_{22}(\boldsymbol{n}) & \rightarrow2-\frac{1}{3}\sum_{i>j}\cos[\frac{2\pi}{3}(n_{i}-n_{j})],\label{Sub1}\\
f_{12}(\boldsymbol{n}) & \rightarrow\frac{2}{3}\sum_{i>j}\cos[\frac{2\pi}{3}(n_{i}-n_{j})]-1\label{Sub2}.
\end{align}
Because, for a set of given $\{n_1,n_2,n_3\}$, this substitution keeps the average of $f_{\mu\nu}(\boldsymbol{n})$ over the permutations of this set invariant. 
Furthermore, we have the summation:
\begin{align}
N^{2}\sum_{n_{1},n_{2}}\frac{N!}{3^{N+1}n_{1}!n_{2}!n_{3}!}=\frac{N^{2}}{3}\label{Sum1},
\end{align}
and
\begin{align}
 & N^{2}\sum_{n_{1},n_{2}}\frac{N!}{3^{N+1}n_{1}!n_{2}!n_{3}!}\frac{1}{3}\sum_{i>j}\cos[\frac{2\pi}{3}(n_{i}-n_{j})]\nonumber\\
= & \frac{N^{2}}{3^{N+1}}\sum_{n_{1}n_{2}}\frac{N!}{n_{1}!n_{2}!n_{3}!}\cos[\frac{2\pi}{3}(n_1-n_2)]\nonumber\\
= & \frac{N^{2}}{3^{N+1}}\sum_{n_{1}n_{2}}\frac{N!}{n_{1}!n_{2}!n_{3}!}\mathrm{Re}\left[(e^{\rm{i}2\pi/3})^{n_{1}}(e^{\rm{i}4\pi/3})^{n_{2}}(e^{\rm{i}0\pi/3})^{n3}\right]\nonumber\\
= & \frac{N^{2}}{3^{N+1}}\mathrm{Re}\left[e^{\rm{i}2\pi/3}+e^{\rm{i}4\pi/2}+e^{\rm{i}0\pi/3}\right]^{N}\nonumber\\
= & 0.\label{Sum2}
\end{align}
Make the substitution (\ref{Sub1}) and (\ref{Sub2}) on Eq.~(\ref{eq:FElem}), then insert the summation (\ref{Sum1}) and (\ref{Sum2}), we have the CFIM entries
\begin{equation}
F_{11}^{c}=F_{22}^{c}=\frac{2N^{2}}{3},F_{12}^{c}=F^{c}_{21}=-\frac{N^{2}}{3}.
\end{equation}
We have thus shown the validity of Eq.~(\ref{Ent}) for particle number $N=3k$ with $k\in\mathbb{N}^+$. Its validity for $N=3k+1$ and $N=3k+2$ can be verified with the same methods.

\end{widetext}

\end{document}